 \documentclass[preprint]{aastex}
\usepackage{epstopdf}

\slugcomment{Submitted to the Astronomical Journal}

\shorttitle{Disappearance of Ly $\alpha$ Blobs}
\shortauthors{Keel et al.}

\begin{document}


\title{The Disappearance of Lyman $\alpha$ Blobs: a GALEX Search at $z=0.8$\footnote{Based on observations made with the NASA Galaxy Evolution Explorer. 
GALEX is operated for NASA by the California Institute of Technology under NASA contract NAS5-98034.}}

\author{William C. Keel
\email{wkeel@ua.edu} 
and Raymond E. White, III}
\affil{Department of Physics and Astronomy, University of Alabama,
Box 870324, Tuscaloosa, AL 35487}

\author{Scott Chapman}
\affil{Institute of Astronomy, University of Cambridge, Madingley Road, Cambridge, CB3 0HA, United Kingdom}

\and

\author{Rogier A. Windhorst}
\affil{School of Earth and Space Exploration, Arizona State University, Tempe, AZ 85281-1404}

\begin{abstract}
Lyman $\alpha$ blobs --- luminous, spatially extended emission-line nebulae, often lacking bright continuum counterparts --- are common in dense environments at high redshift. Until recently, atmospheric absorption and filter technology have limited our knowledge of any similar objects at $z\leq 2$. We use GALEX slitless spectroscopy to search for similar objects in the rich environments of two known cluster and supercluster fields at $z=0.8$, where the instrumental sensitivity peaks. The regions around Cl 1054-0321 and Cl 0023+0423 were each observed in slitless-spectrum mode for 10--19 ksec, with accompanying direct images of 3--6 ksec to assist in recognizing continuum sources. Using several detection techniques, we find no resolved Lyman $\alpha$ emitters to a flux limit of $(1.5-9) \times 10^{-15}$ erg cm$^{-2}$ s$^{-1}$, on size scales of 5--30 arcseconds. This corresponds to line luminosities  of $(0.5-3) \times 10^{43}$ erg s$^{-1}$ for linear scales 35--200 kpc.  Comparison with both blind  and targeted surveys at higher redshifts indicates that the population must have evolved in comoving density at least as strongly as $(1+z)^{3}$.
These results suggest that the population of Lyman $\alpha$ blobs is specific to the the high-redshift Universe.
\end{abstract}

\keywords{galaxies: evolution --- galaxies: clusters: individual (Cl 0023+0423, MS 1054-0321) --- ultraviolet: galaxies  }

\section{Introduction}

The wholesale opening of the Universe at $z>2$ to the study of galaxy populations has invigorated the study of galaxy evolution. Lyman break galaxies, submillimeter and mid-IR selection, and extremely red objects (EROs) trace rich populations whose relations to each other and to galaxies seen here and now are still poorly understood. Perhaps even less well understood are the so-called "''Lyman $\alpha$ blobs". These are luminous and very extended emission nebulae which are known at $z \geq 2.3$, where the redshift brings Lyman $\alpha$ to wavelengths easily imaged with large ground-based telescopes. They have emission-line luminosities as high as $10^{44}$ erg s$^{-1}$ 
and detected sizes exceeding 100 kpc (e.g., Steidel et al. 2000, Keel et al. 1999, Francis et al. 2001). In these respects, they are comparable to the emission-line structures seen around some powerful radio galaxies (as recently reviewed by Villar-Mart{\'{\i}}n 2007, noting roles for both gas infall and jet interactions), but their power sources are much less clear. The field in SSA22 contains a significant overdensity of Lyman-break galaxies as identified by Steidel et al. (2000), and has proven especially  fruitful. A deep narrow-band study using Subaru by Matsuda et al. (2004) identified 35 well-resolved Lyman $\alpha$ emitters in this one structure at $z \approx 3.1$, allowing a detailed census of their properties and distribution relative to Lyman-break galaxies. These luminous, extended emitters also seem to be distinct from the population of much smaller, less luminous Lyman $\alpha$ emitters (e.g., Pascarelle et al. 1996, Cowie \& Hu 1998, Steidel et al. 2000), which are consistent with having young stellar populations and winds (perhaps at quite low metallicities; Keel et al. 2002). The deep Subaru and VLT surveys by Matsuda et al. (2004) and Overzier et al. (2008), respectively, suggest that both these populations trace overdensities in the overall galaxy distribution,
  
A recurring pattern is that Lyman $\alpha$ blobs trace the densest regions of the Universe at these redshifts. They were originally found  in regions otherwise known to be rich in other classes of objects (Keel et al. 1999, Steidel et al. 2000), and conversely, Lyman $\alpha$ blobs identified in deep fields are surrounded by overdensities traced by other populations, such as Lyman-break galaxies  (Prescott et al. 2008). Likewise, redshift-stepped searches in random fields show only a sparse and low-luminosity population of extended emitters outside of dense regions (Saito et al. 2006, Yang et al. 2009).
 
The associated continuum objects are diverse. Some of these structures are centered on AGN (Keel et al. 1999), some on Lyman-break galaxies, some on red galaxies best detected in the near-IR (Francis et al. 2001, Palunas et al. 2004), and some have undetectably faint optical continuum sources (Matsuda et al. 2004, Nilsson et al. 2006). Some of them are associated with submillimeter sources (Chapman et al. 2004, Geach et al. 2005), implying significant dust re-radiation within the objects. Careful analysis of Chandra data shows that, while a large fraction of the most powerful blobs contain AGN, these are often energetically insufficient to power the Lyman $\alpha$ emission (White et al. 2004, 2009; Geach et al. 2009). The blobs centered on AGN appear to be similar to the giant Lyman $\alpha$ nebulae seen around many powerful radio galaxies at $z>2$ (e.g., Heckman et al. 1991, van Ojik et al. 1996, Christensen et al. 2006). The blobs without luminous central objects might require a different ionization source.
 
Various mechanisms have been discussed to account for the origin and luminosities of these objects. In view of the high redshifts where this phenomenon was identified, one particularly intriguing possibility is that we see these objects by cooling radiation during a rapid stage in galaxy assembly --- the accretion of large masses of (more or less) pristine gas. This has been inferred in some cases from multiwavelength data which limit embedded AGN or ongoing star formation to levels far below that needed to power the observed emission (Smith \& Jarvis 2007, Smith et al. 2008). 
Another possibility is photoionization by either AGN or starbursts, but this seems difficult to sustain on energetic grounds: 
optical spectra and X-ray detections show that active nuclei are common, but energetically insufficient to ionize the surrounding gas without contrived geometries or very strong obscuration (so that most of the gas sees a level of continuum radiation that we do not). Furthermore, detection of strong submillimeter emission in some cases indicates masses of dust which are not compatible with the amount of resonant scattering expected within such extensive nebulae; Lyman $\alpha$ would be almost completely absorbed if powered from a small central region (Ohyama et al. 2003). Alternatively, winds from either starbursts or AGN may be viable mechanisms, especially if they interact with an ambient medium
and radiate through shock emission. This makes sense with the detection of lines from O and C in some of the extended structures (Keel et al. 2002), and complex kinematics (Ohyama et al. 2006), seen in some instances.

Some extended Lyman $\alpha$ sources are known in the local Universe. Star-forming galaxies have been found to show Lyman $\alpha$ emission on scales of $\approx 10$ kpc (Keel 2005, Hayes et al. 2005, 2007, \"Ostlin et al.. 2008). These structures are generally decoupled from other components of the galaxies, suggesting that scattering is important (and, indirectly, kinematics, which has a controlling role in the escape of Lyman $\alpha$ photons). The escape fraction of Lyman $\alpha$ is always small ($<15$\%).  In the same luminosity range as the high-redshift objects, a handful of local radio galaxies show similar extensive Lyman $\alpha$ emission (Zirm et al. 2009). The power source for these is not always clear; O'Dea et al. (2004) find structural evidence suggesting a role for widespread star formation. Sampling of such objects remains very incomplete, due to the problems of surveying for Lyman $\alpha$ at small redshifts.

The difficulty of performing narrow-band imaging surveys in the ultraviolet has meant that we know almost nothing about the evolution of Lyman $\alpha$ blobs since the epoch corresponding to $z=2$. 
However, the very low background level encountered in the ultraviolet makes space-based slitless
spectroscopy very sensitive in searching for diffuse emission-line objects in comparison to the optical case, allowing us to look for these objects at much lower redshifts. We describe here such a search, using the UV-sensitive GALEX observatory. We find no significant population of Lyman $\alpha$ blobs in dense environments at $z=0.8$, and reinforcing the idea that these objects were specifically inhabitants of the early Universe.

\section{Observations}

\subsection{Clusters observed}

Our strategy was guided by the properties of Lyman $\alpha$ blobs
found at $z > 2.4$, particularly the rich environment at $z=3.1$
in SSA22. This field attracted interest through an overdensity of
Lyman-break galaxies, which motivated a study by Steidel et al.
(2000) revealing two luminous and very extended Lyman $\alpha$
emitters. The most detailed narrow-band census of this region has been a
Subaru survey by Matsuda et al. (2004). They find as many as 35 
well-resolved Lyman $\alpha$ emitters, of which 90\% are within the areas encompassing
the highest surface densities of the (more numerous) unresolved
Lyman $\alpha$ emitters and Lyman-break galaxies. Thus, Lyman $\alpha$
blobs occur in dense structures at high redshift, suggesting that
we should seek later survivors of this population in analogous 
environments. Since dense regions will undergo substantial 
collapse between $z \approx 3$ and $z \approx 0.8$, we consider
the possibility that the most analogous environment at the lower redshift, with respect
to densities of galaxies and intergalactic material, might be the
outer regions of superclusters as well as the clusters themselves.

The highest-redshift window of high sensitivity for Lyman $\alpha$ 
using the GALEX grisms is centered at $z=0.83$, where the line
falls near the peak effective area (at 2470 \AA\ ) of the NUV grism. 
The dense environments of the Lyman $\alpha$ blobs seen at higher
redshifts suggested that we target rich clusters and superclusters
at later times, including their lower-density outskirts to allow
for possible quenching of their production mechanisms by galaxy processing in cluster cores.
Using NED queries, we selected regions with multiple clusters having
spectroscopic redshifts falling within a single 1.2$^\circ$
GALEX field (about 32 Mpc for the WMAP ''year 5" cosmology 
given by Komatsu et al. 2009,  with H$_0 = 71$
km s$^{-1}$ Mpc$^{-1}$ and flat geometry). This selection gave two
pairs or triplets of clusters within spans $\Delta z < 0.03$, both of which
failed the bright-star avoidance criteria for fields centered on the clusters. However, for the triplet including the rich
cluster Cl 0023+0423 at $z=0.83$, acceptable pointings existed with offsets of up to 10 arcminutes from the richest cluster, keeping all three clusters well within the field covered by the dispersed grism light. A similar offset allowed targeting
of the very rich individual cluster MS 1054.3-0321 at $z=0.82$. Both fields have small
extinctions ($A_{2300} < 0.2$ magnitude, following Schlegel et al. 1998). In the WMAP year 5 cosmology,
the angular scales are quite similar at $z=0.83$ (7.6 kpc/arcsecond following Wright 2006)
and $z=3.1$ (7.75 kpc/arcsecond). Table 1 lists the cluster and field center positions for our observations

\subsection{Observations and Data Processing}

The GALEX mission design and data flow have been described by
Morrissey et al. (2007). A 50-cm telescope illuminates two microchannel detectors, separated into near- and far-UV channels by a dichroic beamsplitter. For spectroscopy, a grism is inserted into the beam, with undeviated central wavelength (UDCW) near 1700 \AA\ . The field is circular with diameter 1.2$^\circ$. Observations are obtained during orbital darkness, which means that long integrations are built from individual exposures typically 1500 s in duration. 
The individual grism exposures were culled to eliminate those with poor aspect reconstruction, using the widths of spectra as well as the quality
inspection reports (and in one case the presence of a distinct set of
interfering spectra from an interval of stable but wrong pointing).
The dispersed observations are listed in Table 2, indicating those
used for further analysis. The total useful spectroscopic integration times were 
18904 seconds for Cl 0023+0423 and 10242 seconds for MS 1054-0321.

Our observing strategy was based on having observations with several different
directions of the grism dispersion. This guards against losing a detection
through overlap with a bright-object spectrum, and lets us retrieve both
position and redshift for objects without continuum counterparts, even
if one detection was compromised. This parallels the basic precepts of Pirzkal et al. (2004) in retrieving spectra from HST ACS grism data. Our ideal data set has these dispersion directions (header keyword GRSPA) split among three values equally spaced 120$^\circ$ apart with comparable exposure times in each orientation. This gives high sensitivity at each orientation and reasonable insurance against loss of detections due to overlap with the spectra of bright objects. In contrast to the GALEX pipeline spectral extraction, we are looking for emission-line structures which may not have continuum counterparts. Thus, we need to have at least two detections at different orientations to derive both the wavelength and sky position of the object. Fig. 1 outlines how this strategy provides these data even when one detection is lost to overlap with a brighter object.

To assist in identifying zero-order
images, we also obtained deep direct exposures in both NUV and FUV
bands. Total integration times (in NUV) were 3290 s for  Cl 0023+0423 and 6738 s for MS 1054-0321.

We used several ways of inspecting the grism data to seek emission-line
objects which may not have continuum counterparts. Guided by the
extent of the known high-redshift Lyman $\alpha$ blobs, we searched for objects as
large as 30 arcseconds in extent. 

The simplest detection technique was inspection of the spectral images, after coadding each set of images at the same orientation. This was done after smoothing with FWHM values ranging from 2--21 arcseconds, and likewise after median filtering with window sizes up to 7 arcseconds. No candidates for extended emission-line features were found from either approach.

To reduce the crowding due to the many continuum objects' dispersed light, we performed an approximate continuum subtraction using a 41-pixel median filter along the dispersion of each image. Operationally, we did this by rebinning each spectral image to align the dispersion with columns, performing the median filter on this version of the image, and then rebinning the median-filtered image back to match the original. In this way we could subtract the median-filtered continuum while preserving the pixel-to-pixel statistics of the original sampling. This approach is still limited by small-scale bumps in the system response, but does eliminate most of the continuum.

The most sensitive and tailored approach to detecting ``orphan" emission-line objects incorporates knowledge of the detector geometry and grism dispersion, plus the expected redshift. The pipeline processing grids the data into a tangent-plane projection with 1.0-arcsecond pixels. The dispersion relation for first-order NUV spectra is given by Morrissey et al. (2007), mapping between wavelength and offset $\gamma$  in pixels according to:
$$\gamma =  -882.1 + 0.7936 \lambda
-2.038 \times 10^{-4}  \lambda^2 + 2.456 \times 10^{-8} \lambda^3 $$
for wavelength $\lambda$ in \AA . The data are reconstructed with an undeviated central wavelength (UDCW) formally 1704 \AA  , so the zero-order images do not match the coordinates of objects in the direct images. In fact, the prism part of the grism assembly gives the zero-order images a nontrivial dispersion in the opposite direction as the first-order spectra.
With this dispersion value, and knowing the expected redshift for objects near the targeted clusters, we can shift each dispersed image so that only objects with emission at a specified wavelength appear at their actual celestial coordinates. Coadding images at various orientations with $\sigma$-clipping rejection will omit any pixel values affected by residual continuum from other spectra, in principle leaving only emission lines at the desired redshift. Including the mean dispersion and PSF width of the GALEX system, we applied this technique to source redshifts spaced every $\Delta z = 0.012$ across the range of redshifts represented by spectroscopically confirmed members of each cluster (using references from NED). This took 3 redshift steps for Cl 0023+0423 and 2 steps for MS 1054-0321. We applied this technique to the continuum-subtracted images described above.

Our typical intensity limit for point sources is of order 0.007 Hz in count rate, or flux $1.5 \times 10^{-15}$
erg cm$^{-2}$ s$^{-1}$. We derive this from the count maps, incorporating a peak effective area of 40.8 cm$^2$ (formally 40.2-41.4 cm$^2$ within the wavelength range we consider). Thus, at 2225 \AA\ , a detected count rate of 1 Hz corresponds to a line flux of $2.18 \times 10^{-13}$ erg cm$^{-2}$ s$^{-1}$. 
With the nominal FWHM for the GALEX PSF of 5 arcseconds, the flux
limit for resolved emission-line objects increases linearly with size
scale. Our typical flux limit for a 10-arcsecond FWHM is $3 \times 10^{-15}$ erg cm$^{-2}$ s$^{-1}$,
and at 30 arcseconds it reaches $9 \times 10^{-15}$ erg cm$^{-2}$ s$^{-1}$.

It is not trivial to assign a single detection threshold to our results, since not only do they depend on the spatial scale that we consider, but the threshold also varies depending on where pixels are affected by spectra of relatively bright objects. One end-to-end test for the reality and level of detections is to coadd the spectral images of a single field shifted to align with each other for noncluster redshifts, or with sets of random offsets corresponding to no physical redshift, and taking their nominal detections as the typical false-alarm rate. This suggests that our
flux limits above generally correspond to 5$\sigma$ except in the rare regions of the image stacks where two zero-order images, or spectra of bright
objects on different input images, overlap.

\section{Comparison with higher-redshift surveys: the evolution of Lyman $\alpha$ blobs}

We detect no resolved Lyman $\alpha$ emission objects in either of our GALEX fields. Our key result is that these two dense regions at $z=0.8$ contain none of the luminous Lyman $\alpha$ blobs which are abundant  overdense regions for $z \geq 2.3$. The large difference in Tolman $(1+z)^4$
surface-brightness dimming between the higher redshifts and $z=0.8$ makes our measurement very sensitive --- our luminosity detection thresholds range from $0.5 - 3 \times 10^{43}$ erg s$^{-1}$
on scale of 5--30 arcseconds.  These flux limits would have allowed us to detect essentially {\it all} of the 35 blobs measured by Matsuda et al. (2004) in the SSA22 region at $z=3.1$, with the qualified wording arising because some of their faintest objects skirt our combined size--flux limit. Counterparts of the most luminous of these high-redshift objects would have been conspicuous over a wide redshift range in the GALEX spectroscopic exposures for the Medium-deep Spectroscopic Survey (MSS) and Deep Spectroscopic Survey (DSS; Deharveng et al. 2008), although their automated spectral extractions are based on the positions of continuum detections, so this statement applies only to their original two-dimensional data.

We assess the significance of this lack of Lyman $\alpha$ blobs by comparison with surveys at higher
redshifts. The evolution of clustering as well as of individual galaxies makes it difficult to know just
what environments are comparable between $z \approx 0.8$ and $z > 2.3$, so that we would ideally
like to examine the entire range of environmental density. The best we can do at this point combines the
surveys of overdense regions (most notably the SSA22 structure at $z=3.1$ and the ``blind" surveys by
Saito et al. 2006 and Yang et al. 2009). The work of Yang et al. is particularly important, with four
detections within 4.8 square degrees at $z \approx 2.3$. We follow the compilation in their Fig. 2, adding our GLAEX limit at $z \approx 0.8$ and calculating comoving volumes following Wright (2006). For nonclustered objects, our data sample the redshift range $z=0.64-1.25$ (half-power transmissions for Lyman $\alpha$ ). Thus,
making no overdensity correction for any assumption about clustering of the blob population, 
and taking the total effective solid angle of our search at 0.40 square degrees to account for the area of the field lost due to spectral dispersion, we derive an upper limit to the comoving density $ < 0.5 \times 10^{-6}$ Mpc$^{-3}$ from detecting $<1$ object. Our upper limit lies well below the Yang et al. ``field" density at $z=2.3$, despite our deliberately including  very dense regions at $z=0.8$, and our somewhat better surface-brightness sensitivity (so we could detect $\approx 3$ times as many Lyman $\alpha$  blobs from Matsuda et al.). We adopt the Yang et al. scaling to a common limiting luminosity $> 1.6 \times 10^{53}$ erg s$^{-1}$ between the various high-redshift surveys, which matches our threshold luminosity for an object size 15 arcseconds or less.

Fig. 2 compares comoving densities of Lyman $\alpha$ blobs from these surveys, where we have added our upper limit to those tabulated by Yang et al. (2009). If we take the frequently-used form $\rho \propto (1+z)^{n}$, we derive values $n>3.0$, $n>3.4$ between the GALEX upper limit and the lower error limits for the Yang et al. field survey and the Matsuda et al. ``cluster" value respectively. Thus our data suggest the robust conclusion that the space density of luminous Lyman $\alpha$ blobs has evolved at least as rapidly as $(1+z)^{3}$.

\section{Summary}

We have used GALEX slitless spectroscopy to search for Lyman $\alpha$ blobs
in known rich cluster and supercluster fields at $z=0.8$, where the near-UV grism mode is most sensitive. Such objects are common in dense regions at
redshifts $z=2.3-4$, so a probe at lower redshift allows us to tell whether
these objects are associated only with the early epochs of galaxy assembly, or might have persisted longer in cosmic time. In addition, their central galaxies (if any) could be studied in considerably more detail at the lower redshift.

We found no resolved Lyman $\alpha$ emitters at the cluster redshift (nor, to somewhat higher flux limits, at any other redshift putting the line within the NUV grism's range of high sensitivity). Our 5$\sigma$ detection limits range from $(1.5-9) \times 10^{-15}$ erg cm$^{-2}$ s$^{-1}$, for size scales of 5--30 arcseconds. Using the WMAP year 5 cosmology, this translates into line luminosities of $(0.5 - 3) \times 10^{43}$ erg s$^{-1}$. Hence, we would have easily detected virtually all the Lyman $\alpha$ blobs known at higher redshifts, if they persisted to $z=0.8$ and occupy either dense clusters or their surroundings out to comoving distances of order  15 Mpc. Comparison with both blind  and targeted surveys at higher redshifts indicates that the population of Lyman $\alpha$ blobs must have evolved in comoving density at least as strongly as $(1+z)^{3}$.

Among the proposed explanations for Lyman $\alpha$ blobs --- cooling radiation from collapsing protogalactic clouds, superwinds from massive starbursts,  photoionization by active nuclei which may be obscured or transient --- most of the discussion has implicitly incorporated the idea that Lyman $\alpha$ blobs were largely a phenomenon of the high-redshifty Universe.  Our GALEX observations provide support for this idea through nondetection of such objects in cluster environments at $z \simeq 0.8$.

\acknowledgments

This work was supported by NASA through grant NNG05GE51G. 
GALEX is a NASA Small Explorer launched in 2003 April. We gratefully
acknowledge NASA's support for the construction, operation, and
science analysis for the GALEX mission, developed in cooperation with the
Centre National d'Etudes Spatiales of France and the Korean Ministry
of Science and Education. This research has made use of the NASA/IPAC Extragalactic Database (NED) which is operated by the Jet Propulsion Laboratory, California Institute of Technology, under contract with the National Aeronautics and Space Administration. WCK acknowledges support
during the preparation of this paper from a Dean's Leadership Board 
Faculty Fellowship. We thank Yujin Yang and collaborators for permission to quote their
results even before they appeared in astro-ph.


\clearpage



\plotone{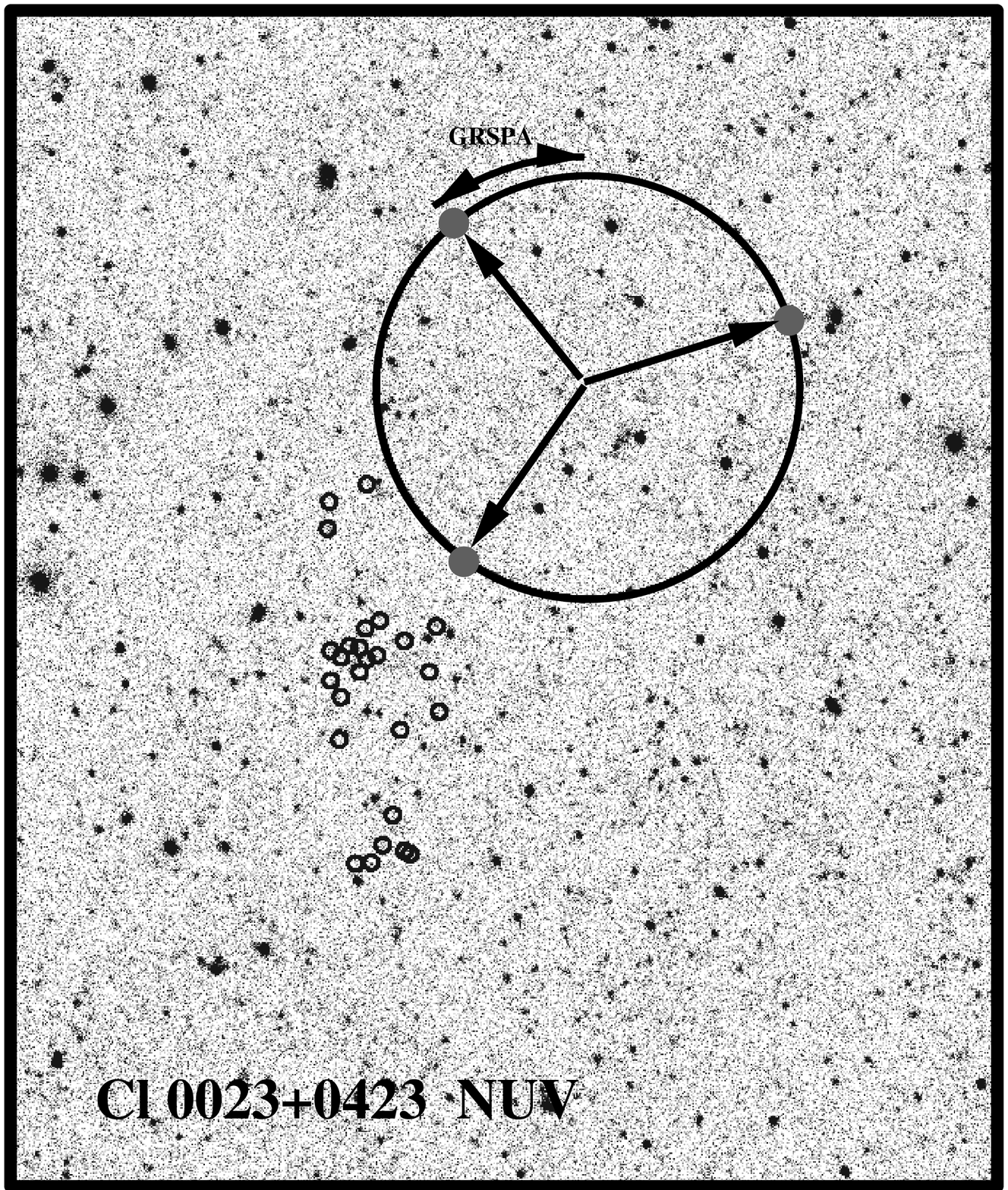}
\figcaption[f1.eps] {A section of the NUV image of the field around Cl 0023+0423 (11.3 by 13.5 arcminutes), illustrating the spectroscopically confirmed cluster members (not detected, shown with 10-arcsecond circles). These indicate the scale of galaxy structures in comparison to the dispersion of Lyman $\alpha$ at the cluster redshift,  and our detection strategy for emission-line objects. The large circle marks possible loci of Lyman $\alpha$ emission from an object at its center as observed at various orientations, with three sample orientations shown. The radius of the circle specifies the object's redshift, and its center specifies its location. The use of at least 3 orientations makes detections robust even if one image is compromised by overlap with another object's continuum.  \label{fig1}}

\includegraphics[scale=1.00,angle=90]{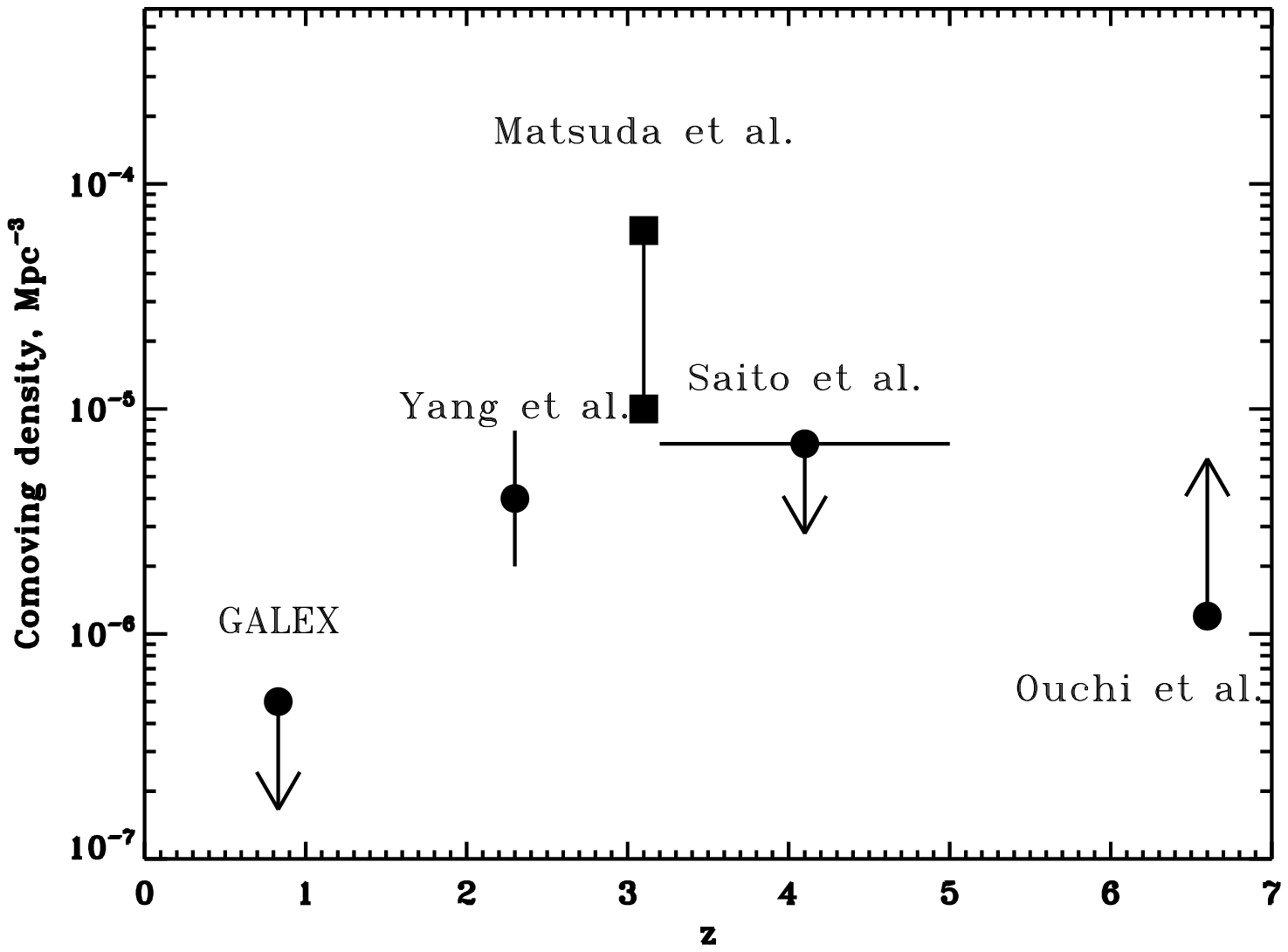}
\figcaption[f2.eps] {Comoving density of Lyman $\alpha$ blobs at various redshifts, including other surveys as summarized by Yang et al. (2009).  The Matsuda et al. (2004) data are shown with and without a correction for the overdensity of the $z=3.1$ structure in SSA22; these points use a square symbol to distinguish them as coming from a targeted (rather than blind) survey, while our data yield even more stringent comparisons if the targeting of rich clusters is relevant to the density of Lyman $\alpha$ blobs. In comparison with Fig. 2 of Yang et al., we have combined their X-ray detections and nondetections.\label{fig2}}

\clearpage

\begin{deluxetable}{lccccc}
\tabletypesize{\scriptsize}
\tablecaption{Cluster regions observed \label{tbl-1}}
\tablewidth{0pt}
\tablehead{
\colhead{Cluster} & \colhead{$z$}   & \colhead{Center:}   &
\colhead{$\alpha$ (2000) $\delta$} &
\colhead{GALEX Pointing center:}  & \colhead{$\alpha$ (2000) $\delta$} 
}
\startdata
Cl 0023+0423 & 0.827  & & 00 23 53.9 +04 23 16   & & 00 20 00 +04 31 48 \\
MS 1054-0321 & 0.823 & & 10 57 00.2 --03 37 27   & & 10 57 02 --03 37 00 \\
 \enddata

\end{deluxetable}

\clearpage

\begin{deluxetable}{lccl}
\tabletypesize{\scriptsize}
\tablecaption{GALEX grism observations \label{tbl-2}}
\tablewidth{0pt}
\tablehead{
\colhead{Observation ID} & \colhead{Exposure (s)}   & \colhead{GRSPA$^\circ$}   &
\colhead{Used?}  
}
\startdata
GI1\_075002\_CL0023p0423\_0001 &  1529.35 & 240.121 & good\\
GI1\_075002\_CL0023p0423\_0002  & 1285.55 & 120.020 & good\\
GI1\_075002\_CL0023p0423\_0003  & 1272.25 & 359.835 & good\\
GI1\_075002\_CL0023p0423\_0004   & 232.75 & 119.794\\
GI1\_075002\_CL0023p0423\_0005   & 920.6  & 239.956 & good\\
GI1\_075002\_CL0023p0423\_0006   & 184.15 &   0.203\\
GI1\_075002\_CL0023p0423\_0007  & 1461.45 & 119.823 & good\\
GI1\_075002\_CL0023p0423\_0008  & 1538.5  & 240.023 & good\\
GI1\_075002\_CL0023p0423\_0009  & 1539.05 & 359.853 &  good\\
GI1\_075002\_CL0023p0423\_0010  & 1540.   & 119.822 & good\\
GI1\_075002\_CL0023p0423\_0011  & 1540.  &  239.825 & good\\
GI1\_075002\_CL0023p0423\_0012  & 1558.7 &   66.104 & good\\
GI1\_075002\_CL0023p0423\_0013  & 1560.  &   29.757\\
GI1\_075002\_CL0023p0423\_0014  & 1560.   &  29.758 & good\\
GI1\_075002\_CL0023p0423\_0015  & 1580.   &  91.195 & good\\
GI1\_075002\_CL0023p0423\_0016  & 1580.   &  15.031 & good\\
GI1\_075002\_CL0023p0423\_0017  & 1579.2 &  236.415\\
           \\
GI1\_075001\_CL1054m0321\_0002 & 1651. &  45.130\\
GI1\_075001\_CL1054m0321\_0003 & 1652. &  45.190 & good\\
GI1\_075001\_CL1054m0321\_0004 & 1650. & 164.849\\
GI1\_075001\_CL1054m0321\_0005 &  231. & 164.940\\
GI1\_075001\_CL1054m0321\_0006 & 1644. & 165.075 & good\\
GI1\_075001\_CL1054m0321\_0007 & 1642. & 165.186\\
GI1\_075001\_CL1054m0321\_0008 & 1617.1 & 284.842 & good\\
GI1\_075001\_CL1054m0321\_0009 & 1640.1 & 285.027\\
GI1\_075001\_CL1054m0321\_0010 & 1641. & 285.042 & good\\
GI1\_075001\_CL1054m0321\_0011 & 1640.  & 285.100\\
GI1\_075001\_CL1054m0321\_0012 & 1700.  & 45.154\\
GI1\_075001\_CL1054m0321\_0013 & 1685.25 & 44.999 & good\\
GI1\_075001\_CL1054m0321\_0014 & 1606.  & 45.282 & good\\
GI1\_075001\_CL1054m0321\_0015 &  202.  & 44.939\\
GI1\_075001\_CL1054m0321\_0017& 1299. & 165.015\\
GI1\_075001\_CL1054m0321\_0016 &  397. &  44.999 & good\\ 
\enddata

\end{deluxetable}

\end{document}